\documentclass{article} 
\usepackage{mlgenx2024,times}

\usepackage{amsmath,amsfonts,bm}

\def\eqref#1{equation~\ref{#1}}

\def\1{\bm{1}}

\DeclareMathAlphabet{\mathsfit}{\encodingdefault}{\sfdefault}{m}{sl}
\SetMathAlphabet{\mathsfit}{bold}{\encodingdefault}{\sfdefault}{bx}{n}

\usepackage{hyperref}
\usepackage{url}
\usepackage{microtype}
\usepackage{relsize}
\usepackage{graphicx}
\usepackage{subfigure}
\usepackage{booktabs} 
\usepackage[percent]{overpic}

\usepackage{xcolor}
\usepackage{bm}

\usepackage{amsmath}
\usepackage{amssymb}
\usepackage{mathtools}
\usepackage{amsthm}

\usepackage[capitalize,noabbrev]{cleveref}

\title{Season combinatorial intervention predictions with \textsc{Salt} \& \textsc{Peper}}

\author{Thomas Gaudelet, Alice Del Vecchio, Eli M Carrami, Juliana Cudini \& Chantriolnt-Andreas Kapourani \\
Relation Therapeutics \\
London, UK\\
\texttt{\{thomas, alice, eli, juliana, andreas\}@relationrx.com} \\
\And
Caroline Uhler \\
Laboratory for Information and Decision Systems, MIT  \\
Eric and Wendy Schmidt Center, Broad Institute of MIT and Harvard\\
Boston, USA\\
\texttt{cuhler@mit.edu}
\AND
Lindsay Edwards \\
Relation Therapeutics \\
London, UK \\
\texttt{lindsay@relationrx.com}
}

\iclrfinalcopy 
\begin{document}

\maketitle

\begin{abstract}
Interventions play a pivotal role in the study of complex biological systems. In drug discovery, genetic interventions (such as CRISPR base editing) have become central to both identifying potential therapeutic targets and understanding a drug's mechanism of action. With the advancement of CRISPR and the proliferation of genome-scale analyses such as transcriptomics, a new challenge is to navigate the vast combinatorial space of concurrent genetic interventions. Addressing this, our work concentrates on estimating the effects of pairwise genetic combinations on the cellular transcriptome. We introduce two novel contributions: \textsc{Salt}, a biologically-inspired baseline that posits the mostly additive nature of combination effects, and \textsc{Peper}, a deep learning model that extends \textsc{Salt}'s additive assumption to achieve unprecedented accuracy. Our comprehensive comparison against existing state-of-the-art methods, grounded in diverse metrics, and our out-of-distribution analysis highlight the limitations of current models in realistic settings. This analysis underscores the necessity for improved modelling techniques and data acquisition strategies, paving the way for more effective exploration of genetic intervention effects.
\end{abstract}

\section{Introduction}

Interventions are an essential tool when striving to decode the functioning of complex systems and in particular, to identify causal dependencies (as exemplified by causal learning theory \citep{eberhardt2007interventions}). In the field of biology, researchers routinely use chemical or genetic perturbations to probe and manipulate cellular systems. Notably, interventions are of fundamental importance to drug discovery, where the goal is to design an intervention that can suppress or even reverse disease-related biological processes in targeted and specific ways. Genetic interventions are of special interest as they allow more precise manipulation of individual genes, as opposed to (for example) small molecules that often have multiple cellular targets and wide-ranging effects that are poorly understood \citep{hopkins2008network, lin2017crispr}. Recent advancements in CRISPR technologies (that allow precise editing of genetic sequences) have empowered scientists to unravel the intricate workings of biological systems \citep{dixit2016perturb, frangieh2021multimodal, norman2019exploring, replogle2022mapping}. From a drug discovery standpoint, manipulating specific elements and observing outcomes enables scientists to pinpoint promising hypotheses for the development of new therapeutics \citep{liu2020crispr, chavez2023advances, kim2023crispr} or understand the mode of action of existing ones \citep{jost2018crispr}.

The application of deep learning to the task of estimating intervention effects in biological systems is a relatively recent development \citep{ma2018using, lotfollahi2019scgen, zhang2023identifiability}. With the emergence of large, single-cell resolution datasets, there has been a rapid increase in research activity aiming to model gene expression patterns \citep{lopez2018deep} and how they are impacted by interventions \citep{roohani2023predicting, lotfollahi2023predicting}. However, as elsewhere, uncertainty remains about the best task framing and most appropriate metrics. For example, straightforward performance metrics (such as the RMSE or correlation coefficient between predicted and actual gene expression) often fail to appropriately weight signals that are likely to be of importance to biologists. By contrast, selected researchers have moved to using custom performance metrics (e.g. the error on only the tails of the predicted distributions) that capture more relevant detail. As a modality of interest, the transcriptome has a number of benefits. First, there are established and widely used technologies, such as scRNA-seq, that accurately measure gene expression levels at single cell resolution. Second, transcriptome variation captures low-level cellular states that often proxy higher-level effects, acting as a generic foundation for predictive models designed for specific phenotypic endpoints (this concept -- that a phenotype can be effectively modelled by a gene expression pattern, sometimes referred to as a 'gene signature' -- is widely used in drug discovery and biology). Because of this, we can motivate research and model development around transcriptomics using the same logic as foundation models in machine learning: addressing a low-level, generic task that can be built upon for specific applications. 

Progress in CRISPR technologies that measure interventional impacts on a cell's transcriptome has opened the door to genome-scale analyses in which most / many genes can be intervened upon individually \citep{replogle2022mapping}. Although conducting a genome-wide perturb-seq (the name used for an experiment where all genes are intervened upon, and the transcriptomic effect measured at the single-cell level) screen remains prohibitively expensive at present, it is anticipated that costs will drop considerably in the coming years. The primary challenge moving forward is expected to revolve around understanding the complex effects of \emph{combinatorial interventions} given the vast number of potential combinations (approximately ${2 \times 10^4 \choose n}$, where $n$ is the number of simultaneous perturbations). Further, studying combinatorial interventions becomes critical due to the limitations of single-target approaches in tackling the complexity of biological systems \citep{hopkins2008network}. Combinatorial strategies perturbing multiple genes may help unravel the organisation of complex networks, overcoming redundancies and disrupting adaptive responses. This is perhaps best exemplified by the combinatorial therapeutics leveraged to treat some forms of cancer \citep{shen2017combinatorial, kim2023crispr}.

In this work, we focus only on pairwise genetic combinations due to existing constraints in available public datasets. Additionally, as we expect the cost of genome-scale screens to decrease, we assume that we have already measured the outcomes of targeting each individual gene. Our key contributions are

\begin{itemize}
    \item[(1)] a careful comparison across current state-of-the-art methods using different metrics and grounded with \textsc{Salt}, a biologically-motivated, non-parametric baseline making the assumption that combination effects are mostly additive,
    \item[(2)] \textsc{Peper}, a deep learning model built from the additive inductive bias from \textsc{Salt} and achieving state-of-the-art performances,
    \item[(3)] an out-of-distribution analysis indicating that current solutions are ill-suited to address realistic scenarios and calling for better modelling approaches and data acquisition strategies.
\end{itemize}

\section{Related work}

\paragraph{Context transfer.} The task at hand can be related to the context transfer problem, whereby the effect of an intervention has been measured in some context, for instance a cell line, and we aim to predict the effect it would have in a different context (e.g. a different cell line). This task has been garnering interest from the deep learning community for some years, particularly for transcriptomics response prediction \citep{lotfollahi2019scgen, wu2022variational, lotfollahi2023predicting}. Notably, both context transfer and combinatorial interventions tasks can be addressed in the same framework, as exemplified by CPA \citep{lotfollahi2023predicting}. Conceptually, the subtle difference between the two tasks arise from the \textit{sequential} nature of context transfer, the context precedes the intervention, as opposed to the \textit{parallel} nature of the task at hand, i.e. all targets are affected simultaneously. 

\paragraph{Potential outcomes and counterfactual estimation.} 
A large majority of approaches \citep{lotfollahi2019scgen, wu2022variational, lotfollahi2023predicting} tackling either the context transfer or combinatorial intervention tasks aim to predict what state an input cell, typically a control cell, would have been in had it been treated or intervened upon. These can naturally be viewed under the lens of counterfactual estimation from the potential outcomes causal framework \citep{rubin2005causal}, which is concerned with resolving similar statements in order to estimate causal effects.

\paragraph{Modelling cellular dynamics.} Generating counterfactual predictions would collapse to a simpler task given a complete model of cellular dynamics that can be sampled at will. Despite a long history of this class of models in metabolism (e.g. \citep{edwards2011msb}), dynamic models of the complexity of transcription are rarer \citep{erbe2023transcriptomic,bhaskar2023inferring, ishikawa2023renge}; the lack of large-scale, time-resolved, interventional datasets has limited the development of methods targeting our task of interest through the modelling of cellular dynamics. 

\section{Methods}

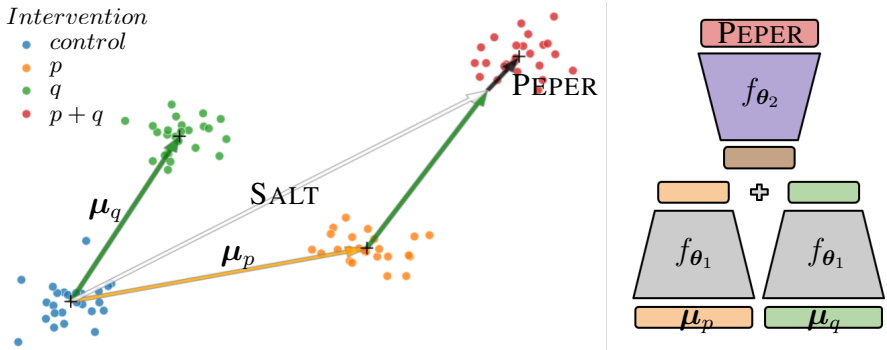
\begin{figure}[t]
\begin{center}
\begin{overpic}[width=0.9\textwidth]{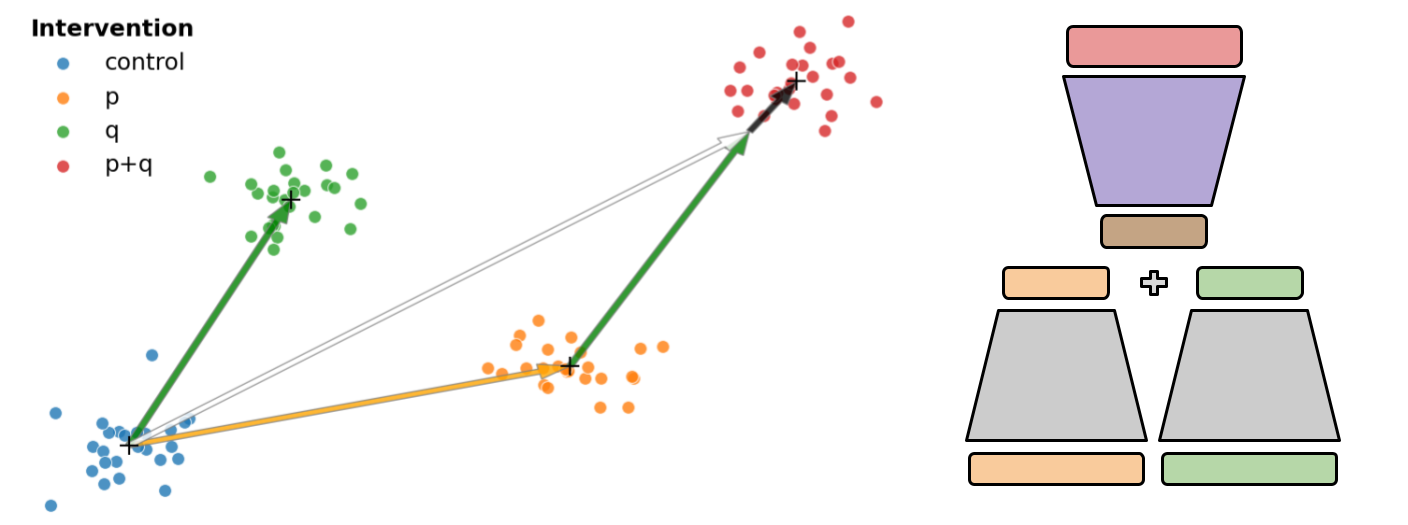}  

 \put (1.5,35) {\colorbox{white}{\small$\displaystyle Intervention$}}
 \put (6.1,32.4) {\colorbox{white}{\small$\displaystyle control$}}
 \put (6.1,29.9) {\colorbox{white}{\small$\displaystyle p$}}
 \put (6.1,27.5) {\colorbox{white}{\small$\displaystyle q$}}
 \put (6.1,24.75) {\colorbox{white}{\small$\displaystyle p+q$}}

 \put (11,15) {\large$\displaystyle\bm{\mu}_q$}
 \put (25,10) {\large$\displaystyle\bm{\mu}_p$}
 \put (28,16) {\large$\displaystyle \textsc{Salt}$}
 \put (55.75,27.5) {\large$\displaystyle \textsc{Peper}$}
 \put (77.5,33) {\large$\displaystyle \textsc{Peper}$}
 \put (87,3.25) {\large$\displaystyle \bm{\mu}_q$}
 \put (73.5,3.25) {\large$\displaystyle \bm{\mu}_p$}
 
 \put (73,10) {\large$\displaystyle f_{\bm{\theta}_1}$}
 \put (87,10) {\large$\displaystyle f_{\bm{\theta}_1}$}
 \put (80,27) {\large$\displaystyle f_{\bm{\theta}_2}$}
\end{overpic}
\caption{Illustration of \textsc{Salt} and \textsc{Peper} predictions for the combinations of interventions on genes $p$ and $q$.}
\label{fig:saltNpeper}
\end{center}
\end{figure}

\subsection{Notation}
In the following, we define $\mathbf{X}_p$ as the matrix representing cells measured after gene p intervention and $\mathbf{\tilde{X}}^m_p$ as the matrix of estimated cells for the same intervention predicted by method $m$, with both matrices structured such that rows correspond to individual cells and columns to measured genes.  We use the letter $c$ to refer to control cells which have only been treated with non-targeting guides and are typically used as controls experimentally. We use lower case $\mathbf{x}$ to denote a single cell, i.e. a row of a matrix as defined previously. If the intervention correspond to a pairwise combination, we write $\mathbf{X}_{p+q}$ where $p$ and $q$ indicate the two genes targeted.

We introduce the \textit{average intervention effect vector} as the average difference between perturbed cells and control cells, we denote it by $\bm{\mu}_p = \overline{\mathbf{X}}_p - \overline{\mathbf{X}}_c$, where $\overline{\mathbf{X}}$ denotes the mean over the columns of $\mathbf{X}$, also known as \textit{pseudo-bulk}, and gives the average expression of each gene across cells. 

\subsection{Existing art}

We use two published baselines to ground our work: GEARS \citep{roohani2023predicting}, the current state-of-the-art, and CPA \citep{lotfollahi2023predicting}. 

CPA follows from a long lineage of autoencoder architectures applied to single-cell (sc)RNA-seq data \citep{lopez2018deep,lotfollahi2019scgen, wu2022variational}. In CPA, the authors model combinatorial effects as an additive process in a learnt (potentially nonlinear) latent space.

By contrast, GEARS takes its inspiration from the causal and graph neural network literature, relying on graph mutilation of biological graph priors and the message passing paradigm. In practice, GEARS applies a fixed, intervention-specific translation to an input cell. 
Formally, given a control cell input (i.e. a cell treated with non-targeting guides), an intervention targeting gene p and q jointly, GEARS's counterfactual prediction can be written as \[\mathbf{\tilde{x}}^{GEARS}_{p+q} = \mathbf{x}_c + g_{\bm{\phi}}(p, q | \mathcal{G}),\] where $g_{\bm{\phi}}$ is a graph neural network architecture with parameters $\bm{\phi}$ and $\mathcal{G}$ denotes the biological graph priors.

\subsection{\textsc{Salt}}

The aim of the study of combinations is often the identification of interactions such as synergy or antagonism \citep{norman2019exploring, bertin2023recover}. However, these interactions are fairly rare \citep{martin2023synergistic} and notably genetic variations tend to be largely additive \citep{hill2008data}. As such, we expect that a simple additive heuristic would provide a strong non-parametric baseline for the task. We refer to it as \textsc{Salt} (Simply Assume Linear combinations of Transcriptomes) and an illustration can be seen on Figure \ref{fig:saltNpeper}. Following our notations, we can write \textsc{Salt} as \[\mathbf{\tilde{x}}^{\textsc{Salt}}_{p+q} = \mathbf{x}_c + \bm{\mu}_p + \bm{\mu}_q.\] It is worth noting that similar vector arithmetic baselines in transcriptome space have been used for the context transfer task \citep{lotfollahi2019scgen}. 

\subsection{\textsc{Peper}}
We spice up \textsc{Salt} by introducing a learnable non-linear correction term. This is reminiscent of statistical models that add a non-interactive term with interactive terms to model and study synergistic effects \citep{norman2019exploring, ronneberg2021bayesynergy}. The resulting method -- Perturbation Effect Prediction by Error Reduction (\textsc{Peper}) -- can be written as 
\begin{align*}
\mathbf{\tilde{x}}^{\textsc{Peper}}_{p+q} = \mathbf{\tilde{x}}^{\textsc{Salt}}_{p+q} + f_{\bm{\theta}}(\bm{\mu}_p , \bm{\mu}_q ), 
\end{align*}
where $f_{\bm{\theta}}$ is a neural network with parameters $\bm{\theta}$.  In practice, we decompose $f_{\bm{\theta}}$ as follows \[f_{\bm{\theta}} (\bm{\mu}_p , \bm{\mu}_q ) = f_{\bm{\theta}_2} ( f_{\bm{\theta}_1}(\bm{\mu}_p ) + f_{\bm{\theta}_1}(\bm{\mu}_q )),\] with both $f_{\bm{\theta}_{1,2}}$ corresponding to multi-layer perceptrons. See Figure \ref{fig:saltNpeper} for illustration. From the definitions, it is important to remark that GEARS, \textsc{Salt} and \textsc{Peper} would at best capture the average intervention effect vector, $\bm{\mu}_{p+q}$. 

To train \textsc{Peper} we use the Central Moment Discrepancy (CMD) which is a distributional loss that compares moments \citep{zellinger2016central}. Because \textsc{Peper} only models the average effect of interventions, we only use the first moment to compute the loss. Note that with this loss, we batch the data by intervention, i.e. each batch contains data from a single intervention. 

\section{Experiments}

\subsection{Evaluation}

\paragraph{Datasets.} We use two distinct CRISPR interventional datasets to evaluate prediction of the outcome of genetic combinations for our experiments \citep{norman2019exploring, wessels2023efficient}. Details can be found in Appendix \ref{data-processing}.

\paragraph{Splits.} For each datasets, we define two in-distribution splits, \textit{in-distribution-25} and \textit{in-distribution-75}, and one out-of-distribution split based on similarities between intervention effects in transcriptomics space, reasoning that the task is rendered more challenging if a model is forced to estimate out-of-distribution transcriptomic responses. We detail the split creation in Appendix \ref{data-processing}.

\paragraph{Hyperparameters selection.} We tune each model on each split, using the validation sets to identify the best configurations. For CPA and GEARS, we use the tuning configurations provided by the authors. For \textsc{Peper}, we provide the hyperparameters and associated ranges considered in Appendix \ref{peper_tuning}. We retrain each model post-tuning with the identified best configurations using $5$ different seeds.

\paragraph{Metrics.} Following recommendations from \citet{ji2023optimal}, we use root mean squared error (RMSE) as our primary metric to evaluate models' predictions (see Appendix \ref{log-fold-change} for details). We also include energy distance scores, a distributional metric, in the Appendix. Following previous work \citep{roohani2023predicting}, we calculate metrics in two ways: for all genes (labelled \textsc{All} in tables) and for the top $20$ genes ranked by absolute effect sizes (labeled \textsc{Top 20 ES} in tables). These sets of $20$ genes are naturally intervention-specific. As noted in the Introduction, metrics computed on a smaller, meaningful set of genes are generally preferred by the community as many lowly expressed genes are noisy and the effect of genetic interventions tend to be observed on a small set of genes \citep{peidli2024scperturb}. We include both for completeness and report metrics computed on the held-out test sets using the 5 models obtained from the different seeds.

\subsection{Results}

\textsc{Peper} achieves state-of-the-art for in-distribution splits, as shown in Tables \ref{tab:25_results} and \ref{tab:75_results}, indicating that our recipe relying on a strong inductive bias motivated by biological observations is an effective approach to the problem. 
The relatively good performances of the \textsc{Salt} heuristics also speak to the quality of the inductive bias. In particular, it is notable that \textsc{Salt} beats both CPA and GEARS in the hardest in-distribution-75 split of the norman-dataset. Our results demonstrate that the inductive bias provided by \textsc{Salt}'s prior is more useful to the task at hand than the priors that GEARS relies on. 

\begin{table}[t]
    \centering
    \caption{RMSE results on the in-distribution-25 splits. The numbers correspond to average score and standard deviation (in parenthesis) across 5 different seeds.}
    \label{tab:25_results}
\begin{center}
\begin{small}
\begin{sc}
    \begin{tabular}{|c|c c|c c|}
        \hline
         & \multicolumn{2}{|c|}{\textbf{norman-dataset}} & \multicolumn{2}{|c|}{\textbf{wessels-dataset}} \\
        \hline
        Model & \textbf{All} & \textbf{Top 20 ES} & \textbf{All} & \textbf{Top 20 ES} \\
        \hline
        \textsc{Peper} & 0.030 (0.002) &  \textbf{0.104 (0.006)} & \textbf{0.034 (0.001)} & \textbf{0.094 (0.002)}\\
        GEARS & \textbf{0.029 (0.001)} & 0.117 (0.0099) & \textbf{ 0.034 (0.001)} & 0.101 (0.004)	\\
        CPA & 0.032 (0.002) & 0.119 (0.008)	 & 0.036 (0.000) & 0.131 (0.007) \\
        \textsc{Salt} & 0.034 & 0.124 & 0.061 & 0.143\\
        \hline
    \end{tabular}
\end{sc}
\end{small}
\end{center}
\end{table}

\begin{table}[t]
    \centering
    \caption{RMSE results on the in-distribution-75 splits. The numbers correspond to average score and standard deviation (in parenthesis) across 5 different seeds. }
    \label{tab:75_results}
\vskip 0.15in
\begin{center}
\begin{small}
\begin{sc}
    \begin{tabular}{|c|c c|c c|}
        \hline
         & \multicolumn{2}{|c|}{\textbf{norman-dataset}} & \multicolumn{2}{|c|}{\textbf{wessels-dataset}} \\
        \hline
        Model & \textbf{All} & \textbf{Top 20 ES} & \textbf{All} & \textbf{Top 20 ES} \\
        \hline
        \textsc{Peper} & \textbf{0.030 (0.000)} & \textbf{0.118 (0.000)} & 0.043 (0.001) & \textbf{0.112 (0.005)} \\
        GEARS & 0.038 (0.001)  & 0.145 (0.011)& \textbf{0.038 (0.002)} & 0.113 (0.005)	 \\
        CPA & 0.032 (0.000) & 0.139 (0.004) & 0.071 (0.004) & 0.164 (0.018)	 \\
        \textsc{Salt} & 0.033 & 0.134 & 0.065 & 0.160 \\
        \hline
    \end{tabular}
\end{sc}
\end{small}
\end{center}
\vskip -0.1in
\end{table}

\begin{figure}[ht]
\vskip 0.2in
\begin{center}
\centerline{\includegraphics[width=0.8\columnwidth]{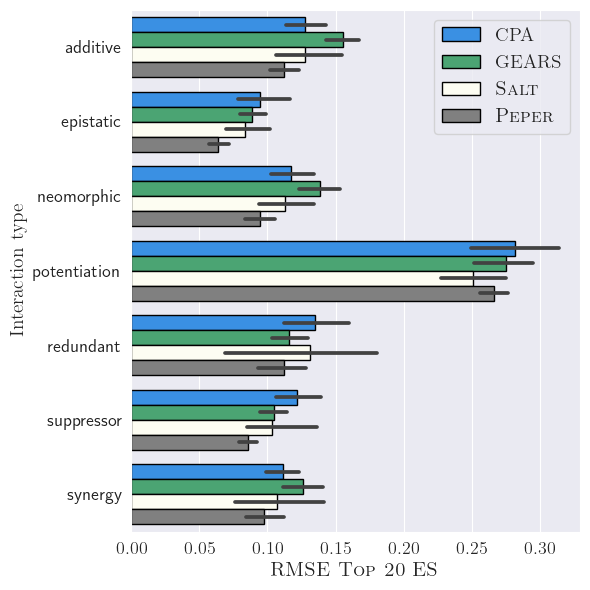}} 
\caption{Score breakdown of GEARS, CPA, \textsc{Salt}, and \textsc{Peper} (lower is better) based on intervention subgroups identified by \citet{norman2019exploring}. The definition for each label can be found in Appendix \ref{glossary}. As we can see, there are some variability in performances based on interaction type, but the potentiation interaction type stands out particularly.}
\label{fig:norman_group_performances}
\end{center}
\vskip -0.2in
\end{figure}

We investigate further model performances on the in-distribution-75 split for the norman-dataset by grouping interventions based on the types of interaction identified by \citet{norman2019exploring}.The definition for the different interaction types can be found in Appendix \ref{glossary}. Overall, we observe similar performances across the different groups, with \textsc{Peper} outperforming all other methods (see Figure \ref{fig:norman_group_performances}). However, the \textit{potentiation} group stands out with significantly worse performances for all models. Note that potentiation occurs when one of the two interventions combined has no effect on its own but enhances the effect of specific interventions when combined. Given this definition, the results could be explained by the fact that the average intervention effect vector is uninformative and does not help predicting the enhancing effect of the intervention. This result exemplifies the limitation of the inductive bias provided by \textsc{Salt}.

It is worth remarking that the only method that actually models single-cell variation, CPA, does not benefit from it. One could argue that this is an artifact of the metric choice as RMSE only compares the first moment of the predicted and actual cell distributions. However, the gap with all three other approaches grows even larger when considering the energy distributional distance as reported in Appendix Tables \ref{tab:ed_25_results} and \ref{tab:ed_75_results}. This might suggests that the noise-to-signal ratio in single-cell data is too high to model higher moments accurately for the task and evaluation we chose.
In other terms, architectures that only model average effects and otherwise conserve the higher moments of the distribution of control cells, such as \textsc{Peper} and GEARS, may be advantaged due to the noisiness of the single-cell data.

When out-of-distribution, all models have significant performance drops and \textsc{Salt} becomes the best performing model for all but one metric (see Table \ref{tab:heldout_clusters_results}). As \textsc{Peper} simply augments \textsc{Salt}, the performance drop is less striking when compared to GEARS. However, the results are symptomatic of an overfit to the seen data and a lack of generalisability beyond the support from the training set.

\begin{table}[t]
    \centering
    \caption{RMSE results on held-out clusters splits. The numbers correspond to average score and standard deviation (in parenthesis) across 5 different seeds.}
    \label{tab:heldout_clusters_results}
\begin{center}
\begin{small}
\begin{sc}
    \begin{tabular}{|c|c c|c c|}
        \hline
         & \multicolumn{2}{|c|}{\textbf{norman-dataset}} & \multicolumn{2}{|c|}{\textbf{wessels-dataset}} \\
        \hline
        Model & \textbf{All} & \textbf{Top 20 ES} & \textbf{All} & \textbf{Top 20 ES} \\
        \hline
        \textsc{Peper}  & 0.032 (0.000) & 0.162 (0.000) & \textbf{0.049} (0.003) & 0.189 (0.005) \\
        GEARS & 0.041 (0.001) & 0.274 (0.01) & 0.053 (0.003) & 0.3 (0.068) \\
        CPA & 0.051 (0.024) & 0.330 (0.142) & 0.070 (0.014) & 0.566 (0.096) \\ 
        \textsc{Salt} & \textbf{0.031} & \textbf{0.157} & 0.065 & \textbf{0.181}   \\
        \hline
    \end{tabular}
\end{sc}
\end{small}
\end{center}
\end{table}

\section{Discussion \& Conclusion}

In this paper, we investigated the estimation of the effect of genetic interventions targeting pairs of genes in diverse data scenarios. We proposed to use a biologically-motivated, non-parametric baseline, that we call \textsc{Salt}, to help put results into perspective. We introduced \textsc{Peper}, a deep learning architecture leveraging \textsc{Salt}'s prior and demonstrated that it achieves state-of-the-art performances on our key metric. 

However, our out-of-distribution splits uncovered significant weaknesses in all current modeling approaches when predicting previously unseen regions of the transcriptional response manifold. Unfortunately, this is precisely the setting in which these models are likely to be the most useful to biologists - predicting previously unseen phenotypic responses to combinatorial perturbations. This indicates a need for further research, accompanied by refined approaches to generating data. Creating situations where we consistently find ourselves within an in-distribution scenario for a given biological question is a challenging endeavour. One promising avenue is through efficient data acquisition strategies, such as active learning \citep{settles2009active, bertin2023recover, scherer2022pyrelational}, which provides a level of control over data generation and experimental design.

Concerning model development, the incorporation of inductive biases derived from prior knowledge, typified by Bayesian learning theory \citep{fortuin2022priors} and here exemplified by \textsc{Peper}, has proven to be a robust approach in many applications. However, the observed performance drops in out-of-distribution splits highlight the limitations with current priors, which may not always be of high quality or suitable for generalization. The insufficiency of the inductive bias from \textsc{Salt} becomes clear from the performance drop over combinations that interact through potentiation. Moreover, the use of priors derived from measurements taken at a single timepoint has inherent limitations. These \textit{static} priors lack insight into any underlying dynamics, and interactions that could arise were sequential measures available. To address these issues and enhance performance on this task, we need both better priors, better data and improved architectures capable of leveraging or rejecting inappropriate prior knowledge.

These challenges can also be indicative of a broader issue related to the scarcity of data, notably around combinatorial interventions. We do not yet know whether gene interactions tend to be conserved across different contexts, facilitating the use of transfer learning approaches to fill in the gap, or whether significant novel data would need to be generated for each new context. In all cases, we expect the situation to improve as the volume of publicly available data increases to a scale of dataset sufficient to train large modern architectures. In the interim, the community should employ clever inductive biases and priors while efficiently acquiring data to address specific questions. 

On a conceptual note, our results suggest that modelling higher moments of single-cell distributions may be detrimental to overall performances on our metrics of interest. This should be caveated by the fact that our observation rests only on a single model, namely CPA. However, it can be further compounded by the fact that \citet{ji2023optimal} identified mean-squared error between pseudo-bulks as the best metric to compare single-cell transcriptomics distributions in multiple applications. Put together, these two results might imply limited utility gain from considering higher moments for selected applications; for these, a more productive focus might be centered on (pseudo-)bulk data. However, this statement needs to be tested in specific cases, and it is important to recognise that it might not hold for many possible downstream applications. Lastly, and perhaps most importantly of all, the choice of metrics should be carefully thought through such that improvement on a chosen metric captures meaningful improvements in practice.

\subsubsection*{Acknowledgments}
This work would not have been possible without the support of the entire Relation Therapeutics team, with special mentions for Edith M. Hessel, Marco P. Licciardello, Mohammad Lotfollahi, Rebecca E. McIntyre, Charles E.S. Roberts, Jake P. Taylor-King,  and John C. Whittaker, for the many insightful discussions.

\bibliography{mlgenx2024}
\bibliographystyle{mlgenx2024}

\appendix

\section{Data}\label{data-processing}

\paragraph{Datasets.} The first dataset was generated with CRISPR activation interventions in K562 cells \citep{norman2019exploring}, and comprise $124$ combinations derived from a pool of $106$ individual potential targets. The second dataset utilises CRISPR Cas13 technology in THP1 cells and encompasses $142$ combinations within a set of $27$ genes \citep{wessels2023efficient}. In the following, we refer to each dataset as norman-dataset and wessels-dataset. We detail our data processing in Appendix \ref{data-processing}.

\paragraph{Pre-processing.} We follow standard pre-processing steps for both datasets. We first remove cells that do not pass the usual quality control thresholds in terms of total number of counts, number of genes expressed, and percentage of mitochondrial counts. We also remove the subset of cells associated to a target if the subset is too small. Then, we filter out measured genes if they are not expressed consistently for at least one of the cell subsets associated to targets. Finally, we normalise cell counts to a $10^4$ total and log-transform the data.

\paragraph{Splits.} To create our splits, we first cluster interventions based on pseudo-bulk profiles using the Leiden algorithm \citep{traag2019louvain}. From this we obtain clusters that group interventions that have relatively similar average responses in terms of transcriptomics. We use the clustering to define two \textit{in-distribution} splits and one \textit{out-of-distribution} split. For the \textit{in-distribution} splits, we define train, validation, and test sets from the set of combinatorial interventions within each cluster. The two splits are defined with different amounts of held combinations, we hold-out either 25\% of combinations (i.e. 12.5\% in val and test sets) or 75\%  (i.e. 37.5\% in val and test sets). Henceforth, we will refer to these splits as \textit{in-distribution-25} and \textit{in-distribution-75}. We define \textit{out-of-distribution} splits by holding out entire clusters of combinations for validation and test, choosing the farthest clusters away from the cluster containing control cells. Note that all cells associated to single interventions are added to the training sets for all splits. Visualisations based on UMAP of the various splits can be found in the Appendix (Figure \ref{fig:normps} and Figure \ref{fig:wesps}).

\section{Log-fold change and RMSE metric} \label{log-fold-change}
As discussed in the main document, we compute the root mean-square error between predicted and actual log-fold change vectors with respect to the control distribution. Specifically, we use Limma's definition of log-fold change \citep{smyth2005limma}, which defines the log-fold change associated to perturbation targeting gene $p$ with respect to the control distribution as \[\bm{\delta}_p = \frac{\bm{\mu}_p}{\textnormal{log}(2)},\] assuming the data is log-transformed in base $10$. This formulation of log-fold change from Limma is motivated by the assumption that gene expression follows a normal distribution in log space, rather than in count space. We then denote for each perturbation $p$ the set of $k$ genes that have the greatest absolute log-fold change value by $\mathcal{S}_{p,k}$. With these definitions, we can simply write our metrics over a set of perturbations $\mathcal{P}$ as
\begin{align*}
\textnormal{RMSE}_\textsc{All}(\mathcal{P}) &= \frac{1}{|\mathcal{P}|}\mathlarger{\mathlarger{\sum}}_{p\in\mathcal{P}}\sqrt{\frac{\sum_{i=1}^N \left(\bm{\delta}_p[i] - \tilde{\bm{\delta}}_p[i]\right)^2}{N}}, \\
\textnormal{RMSE}_{\textsc{Top 20 ES}}(\mathcal{P}) &= \frac{1}{|\mathcal{P}|}\mathlarger{\mathlarger{\sum}}_{p\in\mathcal{P}}\sqrt{\frac{\sum_{i\in\mathcal{S}_{p,20}} \left(\bm{\delta}_p[i] - \tilde{\bm{\delta}}_p[i]\right)^2}{|\mathcal{S}_{p,20}|}},
\end{align*}
where $N$ is the total number of genes post-QC, $|\cdot|$ denotes the cardinal of a set, and $\tilde{\delta}$ denotes predicted log-fold changes.

\section{Additional results.} \label{added_results}

To get a view of the prediction quality in distributional terms, we compute the energy distance between predicted and actual distribution of cells on the test set for both in-distribution splits. The results are reported in Tables \ref{tab:ed_25_results} and \ref{tab:ed_75_results}. The results are mostly consistent with the RMSE scores, apart from CPA which performs significantly worse on this metric. As discussed in the main document, this suggest that modelling single-cell variations is detrimental when compared with methods which conserve the higher moments from the control cell distribution.


\section{\textsc{Peper} hyperparameters tuning configuration.} \label{peper_tuning}

We list below the name and range of the hyperparameters tuned for \textsc{Peper} on each split. 
\begin{itemize}
    \item batch size: $\left[256, 6136\right]$,
    \item batch accumulation: $\left[1, 100\right]$,
    \item learning rate: $\left[10^{-5}, 10^{-3}\right]$,
    \item weight decay: $\left[0, 10^{-4}\right]$,
    \item number of encoder layers $\left(f_{\bm{\theta}_1}\right)$: $\left[1, 4\right]$,
    \item number of decoder layers $\left(f_{\bm{\theta}_2}\right)$: $\left[1, 4\right]$,
    \item latent dimension: $\left[1000, 6500\right]$,
    \item encoder and decoder hidden dimension: $\left[1000, 6500\right]$.
\end{itemize}

\section{Glossary.} \label{glossary}

\paragraph{Epistasis} A combination is epistatic if the effect of targeting one of the genes mask the effect of targeting the other.

\paragraph{Neomorphism} A combination is neomorphic if the resulting effect is atypically new or unexpected given the effects observed from intervening on each gene individually.

\paragraph{Potentiation} Potentiation is a sub-type of synergy in that the effect of the combination is greater than expected. The defining aspect is that one of the target does not have any effect on its own and only enhances the effect of targeting the other.

\paragraph{Redundancy} A combination is redundant when the two targets have similar effect when targeted individually and the combination leads also to a similar output.

\paragraph{Suppression} There is suppression when the combined intervention on two genes attenuates the effect of each gene when individually targeted.

\paragraph{Synergy} A combination is synergistic when the effect of targeting both genes together is greater than expected when compared to the effects of targeting each gene individually. There exist multiple mathematical definition of synergy, each making different assumption about what should be expected, e.g. Bliss independence and Loewe additity \citep{tang2015synergy}. 

\begin{table*}[ht]
    \centering
    \caption{Energy distance results on in-distribution-25 split. The numbers correspond to average score and standard deviation (in parenthesis) across 5 different seeds.}
    \label{tab:ed_25_results}
\begin{center}
\begin{small}
\begin{sc}
    \begin{tabular}{|c|c c|c c|}
        \hline
         & \multicolumn{2}{|c|}{\textbf{norman-dataset}} & \multicolumn{2}{|c|}{\textbf{wessels-dataset}} \\
        \hline
        Model & \textbf{All} & \textbf{Top 20 ES} & \textbf{All} & \textbf{Top 20 ES} \\
        \hline
        \textsc{Peper} & 0.151 (0.013) &  \textbf{0.056 (0.003)} & 0.151 (0.002)	 & \textbf{0.041 (0.001)}\\
        GEARS & \textbf{0.144 (0.009)} & 0.062 (0.007) & \textbf{0.126 (0.003)}	 & \textbf{0.041 (0.002)} \\
        CPA & 1.883 (0.048) & 0.129 (0.012) & 2.365 (0.059)	 & 0.181 (0.010) \\
        \textsc{Salt} & 0.145 & 0.065 & 0.292 & 0.059\\
        \hline
    \end{tabular}
\end{sc}
\end{small}
\end{center}
\end{table*}

\begin{table*}[ht]
    \centering
    \caption{Energy distance results on in-distribution-75 split. The numbers correspond to average score and standard deviation (in parenthesis) across 5 different seeds.}
    \label{tab:ed_75_results}
\begin{center}
\begin{small}
\begin{sc}
    \begin{tabular}{|c|c c|c c|}
        \hline
         & \multicolumn{2}{|c|}{\textbf{norman-dataset}} & \multicolumn{2}{|c|}{\textbf{wessels-dataset}} \\
        \hline
        Model & \textbf{All} & \textbf{Top 20 ES} & \textbf{All} & \textbf{Top 20 ES} \\
        \hline
        \textsc{Peper} & 0.151 (0.001) & \textbf{0.067 (0.000)} & 0.211 (0.006) & 0.050 (0.003) \\
        GEARS & 0.178 (0.005)  & 0.085 (0.007) & \textbf{0.152 (0.009)} & \textbf{0.048 (0.003)} \\
        CPA &  1.873 (0.028) & 0.132 (0.005) & 2.467 (0.133) & 0.147 (0.020)\\
        \textsc{Salt} & \textbf{0.146}& 0.075 & 0.331 & 0.069 \\
        \hline
    \end{tabular}
\end{sc}
\end{small}
\end{center}
\end{table*}

\begin{figure*}[ht]
\begin{center}
\centerline{\includegraphics[width=\textwidth]{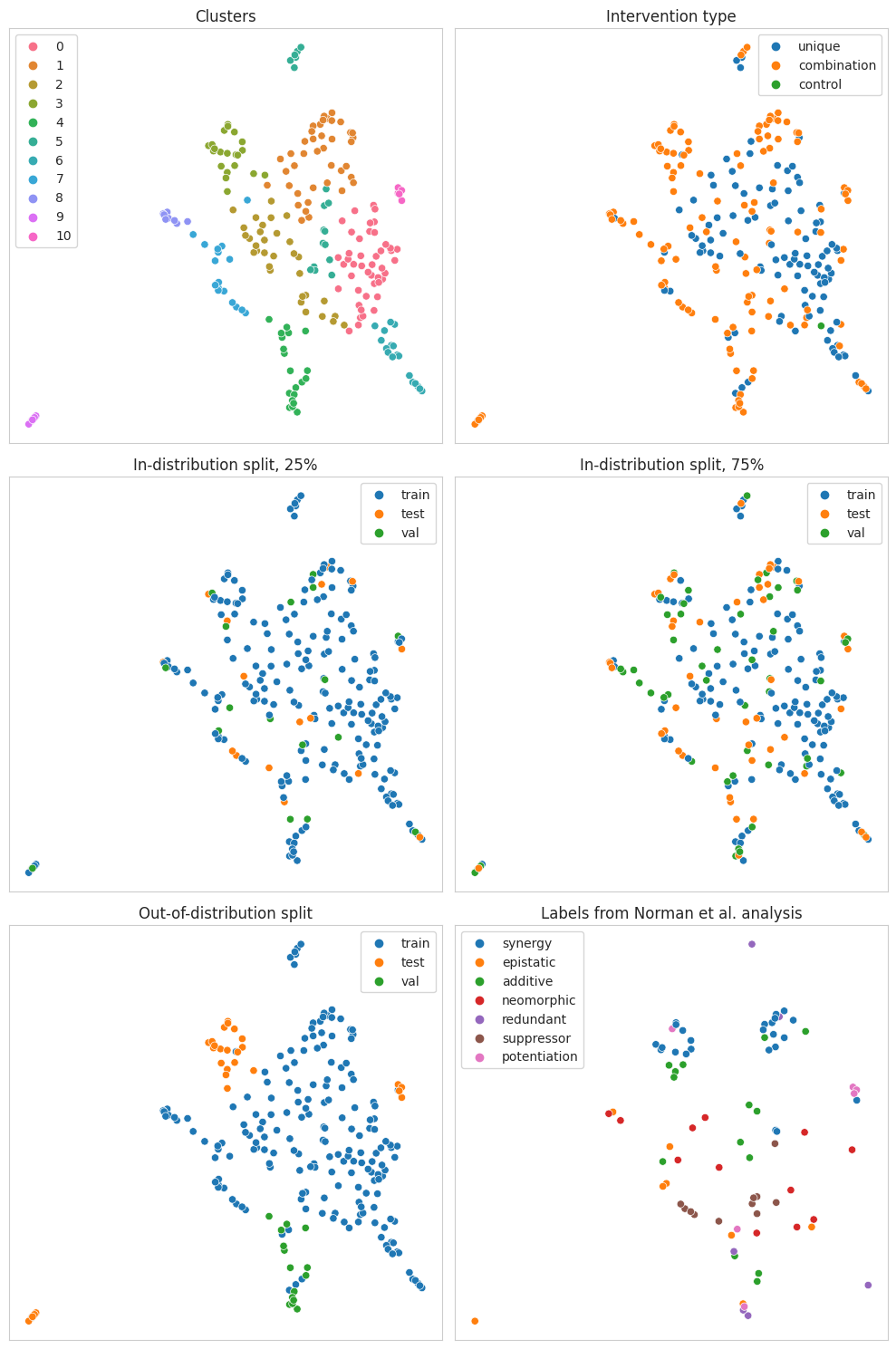}} 
\caption{UMAP visualisation of pseudo-bulked data from the norman-dataset, showing the clustering of interventions, their types, and whether they are in train, validation, or test sets in the different splits we use for the analysis. The last panel show the interaction labels associated to combinations by \citet{norman2019exploring}.}
\label{fig:normps}
\end{center}
\end{figure*}

\begin{figure*}[ht]
\begin{center}
\centerline{\includegraphics[width=\textwidth]{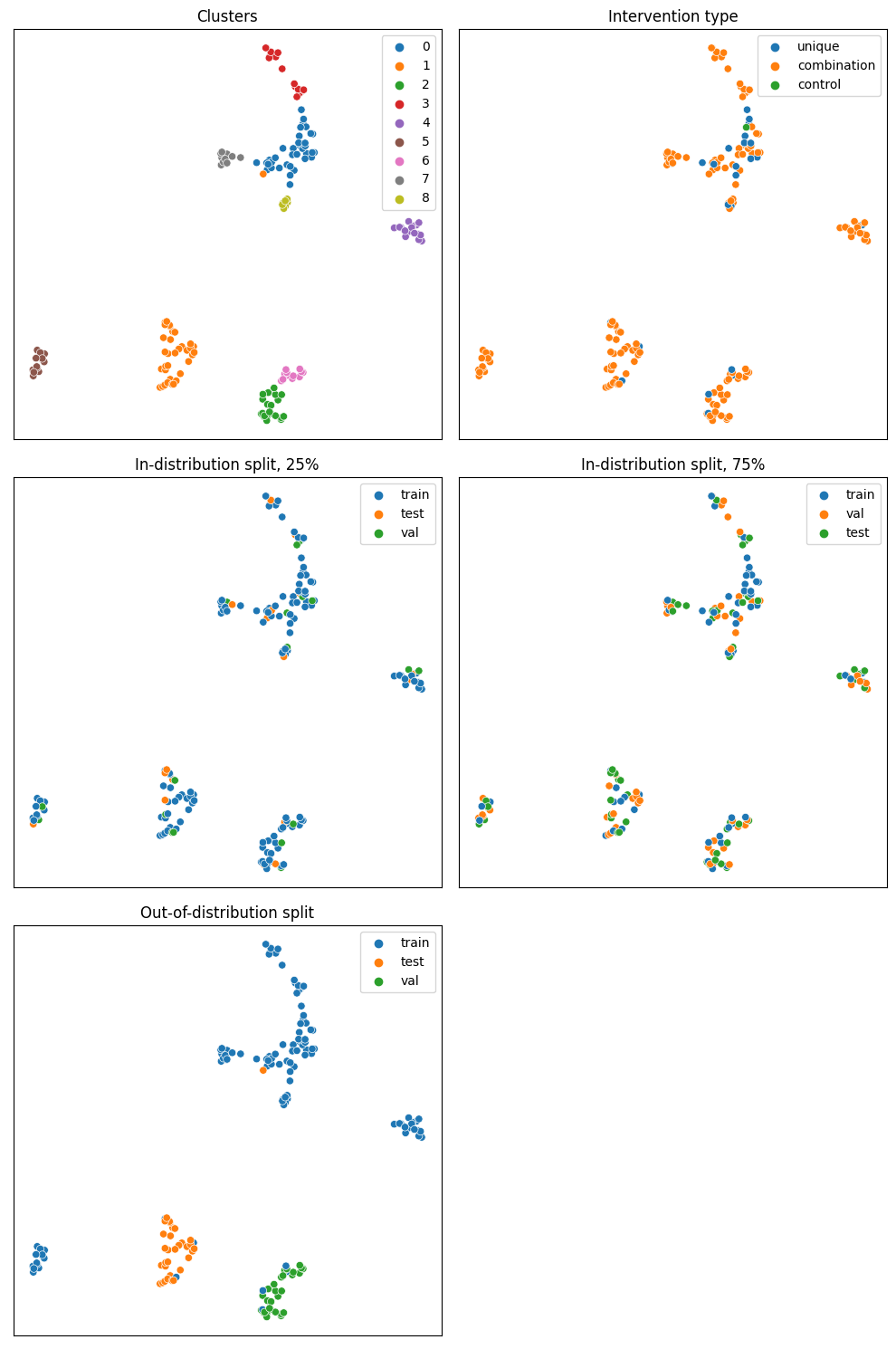}} 
\caption{UMAP visualisation of pseudo-bulked data from the wessels-dataset, showing the clustering of interventions, their types, and whether they are in train, validation, or test sets in the different splits we use for the analysis.}
\label{fig:wesps}
\end{center}
\end{figure*}

\end{document}